\documentclass[conference]{IEEEtran}
\IEEEoverridecommandlockouts
\usepackage{cite}
\usepackage{CJKutf8}
\usepackage[utf8]{inputenc}
\usepackage{amsmath,amssymb,amsfonts}
\usepackage{algorithmic}
\usepackage{graphicx}
\usepackage{textcomp}
\def\BibTeX{{\rm B\kern-.05em{\sc i\kern-.025em b}\kern-.08em
    T\kern-.1667em\lower.7ex\hbox{E}\kern-.125emX}}

\begin{document}

\title{Advance gender prediction tool of first names and its use in analysing gender disparity in Computer Science in the UK, Malaysia and China\\
}

\author{\IEEEauthorblockN{ Hua Zhao}
\IEEEauthorblockA{\textit{School of Mathematical and Computer Sciences} \\
\textit{ Heriot-Watt University}\\
Edinburgh, UK \\
hz103@hw.ac.uk}
\and
\IEEEauthorblockN{Fairouz Kamareddine}
\IEEEauthorblockA{\textit{School of Mathematical and Computer Sciences} \\
\textit{Heriot-Watt University}\\
Edinburgh, UK \\
f.d.kamareddine@hw.ac.uk}
}

\maketitle

\begin{abstract}
 Global gender disparity in science is an unsolved problem. Predicting gender has an important role in analysing the gender gap through online data. We study this problem within the UK, Malaysia and China. We enhance the accuracy of an existing gender prediction tools of names that can predict the sex of Chinese characters and English characters simultaneously and with more precision. During our research, we  found that there is no free gender forecasting tool to predict an arbitrary number of names. We addressed this shortcoming by providing a tool that can predict an arbitrary number of names with free requests.  We demonstrate our tool through a number of experimental results. We show that this tool is better than other gender prediction tools of names for analysing social problems with big data. In our approach, lists of data can be dynamically processed and  the results of the data can be displayed with a dynamic graph. We present experiments of using this tool to analyse the gender disparity in computer science in the UK, Malaysia and China. 
\end{abstract}


\section{Introduction}
In recent years, the problem of global gender disparity in science has occupied an important place amongst governments, academia and companies \cite{b3}. Some researchers have been doing some initial analysis of the situation of the gender gap in academic areas \cite{b3}. Gender prediction methods have been widely used for analysing gender disparities in science on many published articles.  These methods could be enhanced by choosing the most suitable prediction method for a given purpose with optimal parameters and performing validation studies using the finest data source \cite{b12}. In this paper, our purpose is to provide a dynamic tool to analyse the gender gap in computer science in the UK, Malaysia and China. As part of our research, we needed to extend the gender prediction tool for analysing the gender gap in science due to the drawbacks which affect usability in gender disparity studies. More specifically, in the popular existing gender prediction systems, we found that there are no suitable existing systems that can predict a significant number of names for free requests. So, we extended the tool to accommodate an arbitrary number of names for free requests.  Furthermore,  we adapted our tool so that it predicts gender on both Chinese names and English names simultaneously.  
We enhanced the accuracy of our tool so that it performs  better than existing tools. Our implemented tool can be useful for social researchers to analyse large data effectively. Moreover, our tool  can also display the result of the data analysis directly and instantly.

In this paper, we describe our  more accurate gender prediction tool of first names that can predict names on Chinese characters and English characters with big data simultaneously and we use this tool to help analyse the gender disparity in science in the UK, China and Malaysia. 

Our contributions are:
\begin{enumerate}
 \item Enhancing the accuracy of a gender prediction tool for both English and Chinese names simultaneously.
 \item Using the tool in experiments to obtain useful results about gender equality in STEM fields. 
 \item Allowing unlimited free requests when predicting gender with names.
 \item Instantly processing dynamic graphs  as the experiments are run.
\end{enumerate}
 
In section 2, we describe the related work and the reason for improving the system. In section 3, we start with an  existing system that we use as the basis for our extended and generalised tool, and then we describe our new tool in detail.  In section 4, we describe the data for training and testing and for analysing in detail. In section 5, we will outline the experiments' results of testing the system. We will show some results of gender disparity in Computer Science in the UK, Malaysia and China. In section 6, we conclude and give some future work.

\section{Related Work}
There has been much research on doing global gender disparity in science \cite{b3}. Cassidy et al.(2013) \cite{b3} asserted that there might exist a  relationship between  certain disciplines (or cultures)  and the gap of scientists' gender. To continue with their research, we propose to analyse the disciplines and cultures of those scientists. While researching the data, we found that there are many existing gender prediction tools to predict gender by using people's name, such as GenderizeR, Gender API and Ngender \cite{b12, b5,b2}. GenderizeR uses people's first name to predict gender \cite{b12}. However, it can not predict names with Chinese characters. Gender API uses the full name to predict gender and cultural origin \cite{b5}. But it is an online API, and it costs money and is rather costly for an unlimited number of gender prediction. Ngender is a gender prediction tool that can predict Chinese characters, but it does not work with English characters \cite{b2}. In the study of gender disparity in Computer Science, we need to analyse data which contains an arbitrary combination of Chinese characters and English Characters.  Hence, our first task is to create a tool that can predict gender in a file of data with an arbitrary combination of Chinese and English names.  In  our  gender prediction  tool, we use a Naive Bayes classifier for gender prediction:

\subsection{Naive Bayes classifier: Gender Prediction}
The Naive Bayes classifier is a basic classifier \cite{b6}. It uses Bayes Theorem to predict the probability that a given name set belongs to a particular gender, $P(c \mid x)$, from $P(c)$, $P(x)$, and $P(x \mid c)$ 
\cite{b8}. 

The original formula of the Naive Bayes algorithm is as follows:

\[P(c \mid x) = P(c) * P(x \mid c) / P(x).\]

 The existing tool Ngender, uses Naive Bayes classifier based on a suitable formula for gender prediction \cite{b2}:

 \[\begin{array}{l}
 P(gender \mid name) = \\
 \hfill P(gender) * P(name \mid gender) / P(name).
 \end{array}
 \]

In the formula, $P(gender \mid name)$ is the posterior probability of class (gender) given predictor (names); P(gender) is the prior probability of class; $P(name \mid gender)$ is the likelihood which is the probability of predictor given class (gender); P(name) is the prior probability of predictor \cite{b16}. 

\subsection{Existing gender prediction Tools of Names}
Several gender prediction tools of names have been published online. The  five most popular  gender prediction tools are: GenderizeR, Gender API, Ngender, TEXTGAIN and namsor \cite{b12, b5, b2, b11, b14}. These tools can predict genders from people's names and  are used for business and science research.  Table \ref{tab1} shows some information about these tools.  

\medskip

\begin{table}[htbp]
\caption{Existing Gender prediction Tools of Names}
\begin{center}
\hspace*{-5cm}
\begin{minipage}{0.2\textwidth}
\begin{tabular}{|p{1.0cm}|p{1.0cm}|p{1.0cm}|p{1.0cm}|p{1.0cm}|p{1.0cm}|}
\hline
\textbf{Existing Tools} & \textbf{\textit{Genderize R API \cite{b12}}}& \textbf{\textit{Gender API \cite{b14}}}& \textbf{\textit{Ngender \cite{b2}}}& \textbf{\textit{TEXT GAIN \cite{b11}}} & \textbf{\textit{Namsor \cite{b5}}}\\
\hline
\textbf{Language Services}  & 89 Languages & 178 Languages & Chinese &13 Languages &All languages \\
\hline
\textbf{Supported Computing languages }  & R;  Ruby;  Python;  Java;  PHP & PHP;  jQuery;  Java;  Python; PHP legacy & Python & R; Java; JavaScript; PHP; Python;  Ruby; Curl & Android, C\#,  ActionScript,  Java, Objective-C, PHP, Python (v2), Ruby, Scala \\
\hline
\textbf{Service Environment } & Online and Offline & Online  and Offline & Online and Offline  & Online & Online and Offline \\
\hline
\textbf{Reaction } & Limited at 1000 names/day free requests; Can predict few Chinese characters & Limited at 500 names free requests; Can predict limited Chinese characters, but can be  incorrect &  Can only predict Chinese names (Unlimited requests) & 3,000 characters per request (100 free requests per day); It cannot predict Chinese characters & Limited at 1000 names per month free requests; It has errors on predicting Chinese names \\
\hline
\textbf{ The structures of predicting results } &Gender; Probability; Count & Gender; Samples; Accuracy; Duration &  Gender; Probability & Gender; Confidence & Scale; Gender\\
\hline
\textbf{ Requirement of the Input Data for prediction} & Can only input First Names & Can input full Names (cannot identify the first name from Chinese full Names) & Can Input full names in Chinese characters & Can input full Names (does not work for Chinese Names) & Can input Full names (but before input, user needs to classify the full names into First name and Surname)\\
\hline
\end{tabular}
\label{tab1}
\end{minipage}
\end{center}
\end{table}

Some existing gender prediction systems of names can predict lists of Chinese names and English names, (e,g.NamSor, Genderize API, Text Gain and Gender API) \cite{b5, b12, b14, b11}. NamSor only can predict 1000 names per month for free requests \cite{b5}. We tested NamSor and found that some Chinese names cannot be identified. This problem also happens on Gender API \cite{b14}. In NamSor, users have to classify all the names into first name and Surname before they input names for prediction.  Text Gain can predict names when users input original data documents. For example, a  user can predict gender with a  CSV file. However, this function in the system does not work when we tested it with our real data \cite{b11}. Text Gain can predict Chinese names in PinYin \cite{b11}. However, there are lots of Chinese names that have the same characters in PinYin and in such case, PinYin cannot identify the gender of names with a high accuracy. We also found that Genderize R API has the same situation in that it can predict Chinese PinYin and only can predict few Chinese names in characters \cite{b12}.  Genderize R API  can only identiy the first names for predicting genders \cite{b12}. Gender API can predict full names, and when the user inputs original data (e,g. a list of full names), this system is able to classify it into first name and surname.  However, it can not identify Chinese names with Chinese characters \cite{b14}. The  gender prediction tool of names that can more comprehensively predict Chinese names, is Ngender. However, Ngender can not predict English names \cite{b2}. 

In this paper, we aim to predict a large list of names with genders in English names and Chinese names with three datasets. They are the data of people who published papers in the UK, China and Malaysia in Computer Science.  However, the above mentioned gender prediction systems can not help us to predict these datasets directly. Therefore, we  implemented a new tool that can predict any number of combinations of English and Chinese names. This implemented system tool  will be explained in next section.

\section{Implemented System}

\begin{figure}[htbp]
\caption{Basic functions from Existing tools, novel functions from implemented system}
\includegraphics[width=0.5\textwidth]{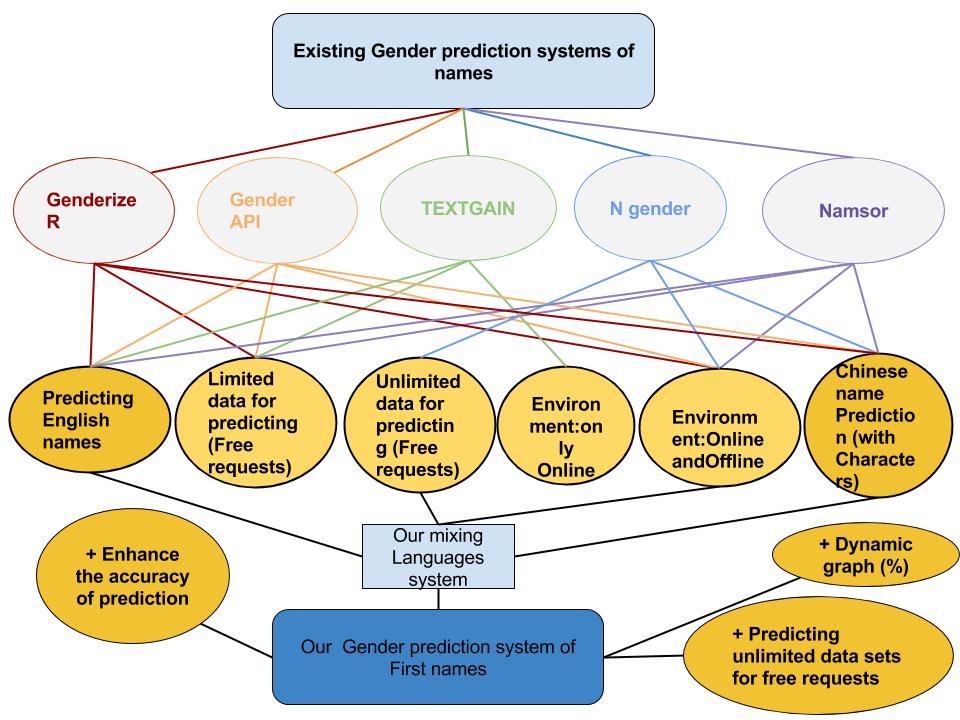}
\label{fig:Namepredictionsystem}
\end{figure}

\begin{figure}[htbp]
  \caption{Gender Prediction Accuracy on existing systems and our Tool}
    \centering
    \includegraphics[width=0.5\textwidth]{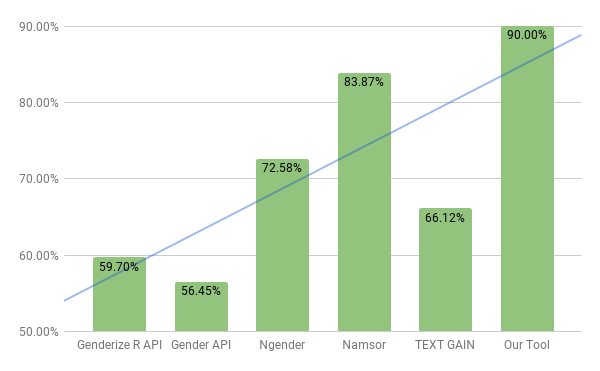}
    \label{fig:accuracychart}
\end{figure}

In this section, we will describe how we implemented an extension of  a popular existing gender prediction system of names, Ngender \cite{b2}. In section 2, we described some information of this existing tool. Figure \ref{fig:Namepredictionsystem} displays the main functions of five existing systems and the improvement of our tool  compared to the existing tools.  
The advantage of our tool is that we enhance the accuracy of the gender prediction in these six systems. Figure \ref{fig:accuracychart} shows the percentage accuracy of our tool and the other five existing tools.  We used 61 real data names to test with all the tools. They contain Chinese names and English names. These data are collected from Baidu and Wiki. Our tool has the highest accuracy of predicting mixed languages in Chinese names and English names.  In this section we will also  describe how we increased the accuracy of prediction. We also improved our tool so that it can process dynamic graphs simultaneously as the experiments are run. The next advantage of our tool is that it can predict unlimited data sets for free requests. 

\begin{figure}[ht]
 \caption{Ngender  and our Tool}
    \centering
    \includegraphics[width=0.5\textwidth]{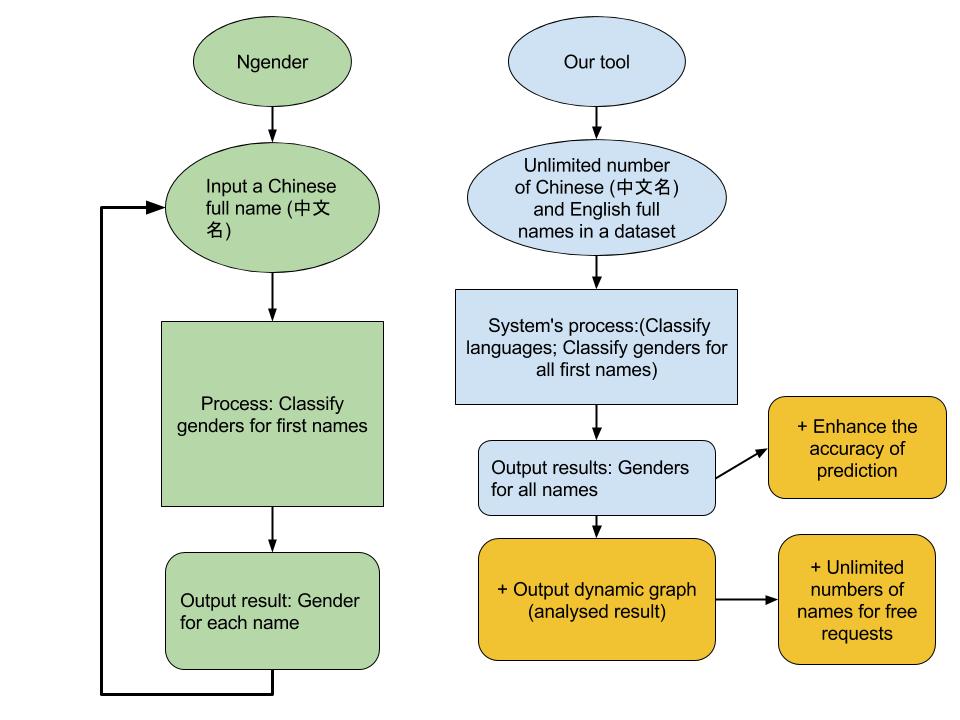}
    \label{fig:comparesystem}
\end{figure}

Figure \ref{fig:comparesystem} shows the difference between our system and the existing gender prediction, Ngender \cite{b2}.
 
\subsection{The functions of the Implemented system}
\begin{figure}[htbp]
  \caption{Window for User - One}
    \centering
    \includegraphics[width=0.25\textwidth]{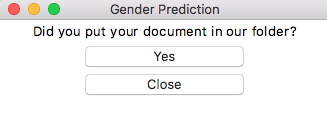}
    \label{fig:gui1}
\end{figure}

\begin{figure}[ht]
   \caption{Window for User -Two}
    \centering
    \includegraphics[width=0.25\textwidth]{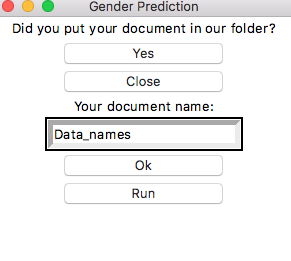}
    \label{fig:gui2}
\end{figure}

\begin{figure}[htbp]
  \caption{ Dynamic graph on analysing the result}
    \centering
    \includegraphics[width=0.5\textwidth]{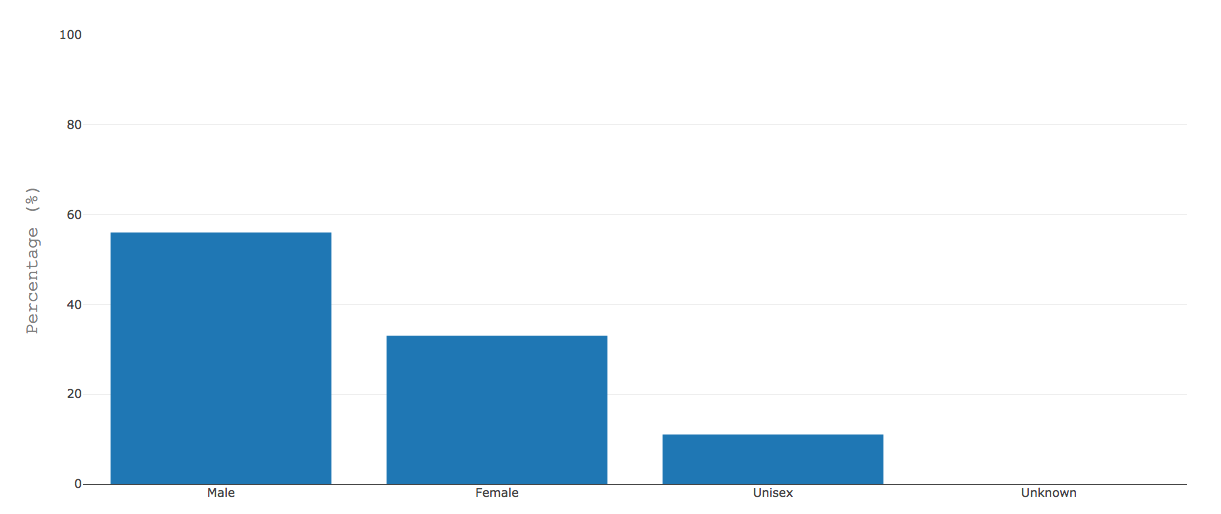}
    \label{fig:resulttest}
\end{figure}

On running our system, the user is informed to put their documents in the folder of the system, see figure \ref{fig:gui1}. Here, the user can input text files and CSV files to predict names.  After the users input their documents, they can input the name of the document they wish to process for predicting, see figure \ref{fig:gui2}. Our system can identify and classify all the names in Chinese and English. After the system processes all the names, it can package a document of the prediction results on all the names. And deliver it to the user's computer. Table \ref{table:Input_res}
shows an example of the input names. Table \ref{table:output_res} displays the results of these names from our system. Our system can classify the genders in male, female and unisex for all the names. After this process, the user can select to get a dynamic graph of this result. Figure \ref{fig:resulttest} shows the result of the example names. For generating the graph, we use a percentage algorithm to predict results in four types of gender classification (Female, Male, Unisex, Unknown). Table \ref{table:classifyItems} shows the definition of the gender classification. For the definition of Unisex, we select the results between 50 \% and 60 \% percentage of each name in Naive Bayes \cite{b15,b2}.  It is also  a method for enhancing the accuracy of gender prediction.  On enhancing the accuracy of gender prediction, our system can classify the original full names in Chinese and English into first names and surnames. This can be more friendly for users since that they do not need to do classification for all the original names. 
For displaying the dynamic graph, we use Plotly Python Library to display the dynamic results \cite{b4}.  

\begin{CJK*}{UTF8}{gbsn}

\begin{table}[htbp]
\centering
 \caption{Input a list of names}
\begin{tabular}{ |p{1cm}||p{3cm}|}
 \hline
Item & Name \\
 \hline
1 & Fairouz Kamareddine \\
\hline
2 & Hua Zhao \\
\hline
3 & Alasdair J G Gray\\
\hline
4 & Phil Barker\\
\hline
5 & Lilia Georgieva \\
\hline
6 & 赵骅 \\
\hline
7 &  赵金标 \\
\hline
8 & 王青 \\
\hline
9 & Jim Thomson \\
\hline
10 & Martin Kettle\\
\hline
 \end{tabular}

\label{table:Input_res}
\end{table}

\begin{table}[!htbp]
\centering
 \caption{Output}
\begin{tabular}{ |p{2cm}||p{3cm}|p{2cm}|  }
 \hline
Item & Name & Gender \\
 \hline
1 & Fairouz Kamareddine & Female\\
\hline
2 & Hua Zhao & Female\\
\hline
3 & Alasdair J G Gray & Male\\
\hline
4 & Phil Barker & Male\\
\hline
5 & Lilia Georgieva & Female \\
\hline
6 & 赵骅& Male \\
\hline
7 & 赵金标 & Male  \\
\hline
8 & 王青& Unisex  \\
\hline
9 & Jim Thomson & Male \\
\hline
10 & Martin Kettle  & Male \\
\hline
 \end{tabular}

\label{table:output_res}
\end{table}
\end{CJK*}

\begin{table}[htbp]
\centering
 \caption{The definition of the gender classification}
\begin{tabular}{ |p{1.5cm}||p{1cm}|p{1cm}|p{1cm}|p{1cm}|}
 \hline
Gender Classification & Female & Male & Unisex & Unkown \\
 \hline
 Percentage & $>$ 60 \% & $>$ 60 \% &  $<$ 50\% AND $>$ 60 \% & None\\
\hline
\end{tabular}

\label{table:classifyItems}
\end{table}

\subsection{Properties of the implemented system}

Our system can predict gender with unlimited numbers of data in Chinese names and English names. For classification and identification of Chinese and English, we use a Python package guess language to identify languages of the input names \cite{b10}. For example, if the system gets the information that this name is "zh " that means it is a Chinese name. When the system identifies the input name is a Chinese name, it can process this name with the Chinese training database to get the percentage number in gender with the first name.  Our system can work with the unlimited datasets for free requests as our system can identify mixed languages in Chinese and English.   The system can output a list of results in one go.  For improving the efficiency of the system, we used a module pickle to process large data and  increase the efficiency of the system \cite{b7}. Table \ref{table:testefficiency} shows the efficiency of our system being testes on different numbers of data.

\begin{table}[htbp]
 \caption{Testing the efficiency of our system on processing data}
\centering
\begin{tabular}{|p{2cm}||p{1.5cm}|p{1.5cm}|}
 \hline
Languages & Number of testing items & Time (Seconds) \\
 \hline
English & 1337 & 1.02135\\
\hline
 Chinese and English & 2901 & 2.26482 \\
\hline
 Chinese and English & 33257 & 22.42112 \\
\hline
 Chinese and English & 133031 & 91.29728\\
\hline
\end{tabular}
\label{table:testefficiency}
\end{table}

\section{Data}

\subsection{Training Data in English Names}

\begin{figure}[ht]
 \caption{The structure of the database for English character in gender prediction tool}
    \centering
    \includegraphics[width=0.4\textwidth]{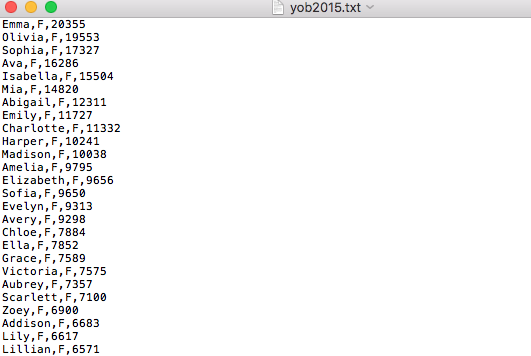}
    \label{fig:databasestructure}
\end{figure}

We collected the data for predicting English characters to improve the gender prediction tool.  THe data is from the National Data on the relative frequency of given names in the population of U.S. births where the individual has a Social Security Number \cite{b9}. The recorded data is collected from the year 1880 to the year 2015 \cite{b9}. 
Figure \ref{fig:databasestructure} shows the structure of the database. In each database, the first column is the name. The second column is the gender of each name, and the third column is the frequency of people used to this name. 

We processed 270 databases, consisting of 95025 names of which 39727 names are male, and 65658 names are female. We cleaned these databases to build one database for all the names and their frequencies of male and female. We used this database as a training database for our system to work with Naive Bayes when predicting genders in English names \cite{b2}.

\subsection{Training Data in Chinese names}

After we cleaned out a feature database of English characters for the system, we collected a database of Chinese characters from Ngender \cite{b2}. This database has the names of Chinese characters and their frequencies. We used this training database for predicting Chinese characters.

\subsection{Testing the accuracy of gender prediction}

For testing the system, we collected data from two websites, Wiki and Baidu \cite{b17,b18}.  The data consists  of the names of famous scientists' names and their genders in the UK and China. There are 162 names of British researchers'  and 122 names of Chinese Scientists. 

\subsection{Data for analysing Gender disparity in Computer Science }

In next section, we will show some results for researching the gender disparity in Computer Science in the UK, Malaysia and China. We collected data from two websites, Thomson Reuters' Web of Science database and CNKI (China National Knowledge Infrastructure) for analysing the gender disparity in computer science \cite{b13,b1}. The data is about the information of articles in Computer Science in the UK, Malaysia and China from 2012 to 2017. 

\section{Experiments}
\subsection{Testing the system}
\begin{figure}[ht]
    \centering
       \caption{Precision of testing the gender prediction system}
    \includegraphics[width=0.5\textwidth]{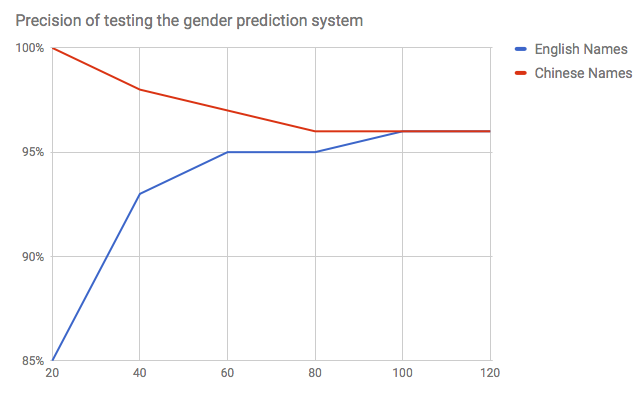}
    \label{fig:TestSystem}
\end{figure}

We tested our system with real collected  data  \cite{b17,b18}. Figure \ref{fig:TestSystem} shows the accuracy of our system. We used 284 researchers names to test our system. There are 162 scientists from the UK and 122 scientists from China.  We know the information of genders from these names. Then we used our system to predict these names' genders. So we compared the results from our tool and the real information to get the accuracy of our system. The accuracy of our system is 96.5 percent.
 
\subsection{Predicting real data of names in analyzing gender disparity in Computer Science}

\begin{figure}[ht]
    \centering
     \caption{Chinese researchers in Computer Science}
    \includegraphics[width=0.5\textwidth]{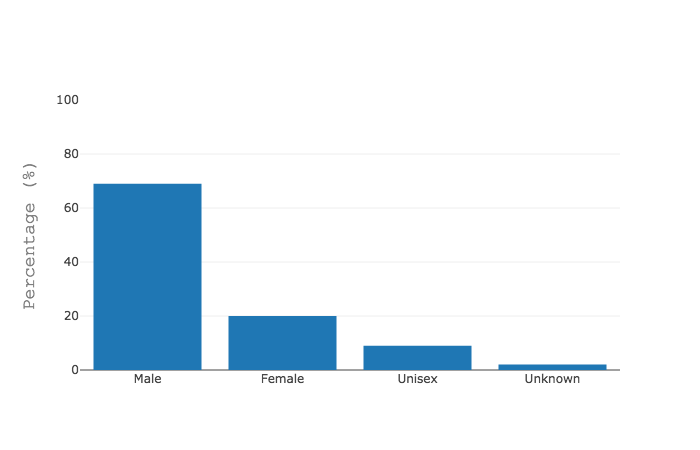}
    \label{fig:CN}
\end{figure}

For analysing the gender gap in computer science, we focused on analysing the places of the UK, Malaysia and China. We used real data from Web of Science and CNKI (China National Knowledge Infrastructure) to analyse the situation in Computer Science to test our system \cite{b1,b13}. Figure \ref{fig:CN} shows the results on the situation of gender disparity in China from 2012 to 2017. We found that more than half of the computing researchers are male.

\begin{figure}[ht]
    \centering
        \caption{UK researchers in Computer Science}
    \includegraphics[width=0.5\textwidth]{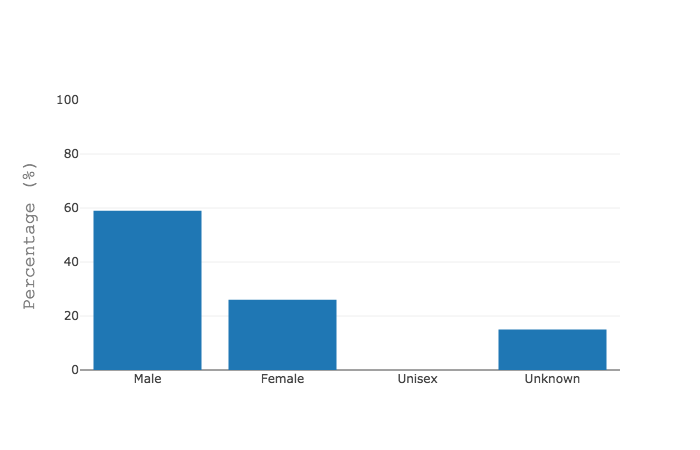}
    \label{fig:UK}
\end{figure}

We also found that the situation is similar in the UK that more than half of male is the computing researchers. Figure \ref{fig:UK} shows the result of the situation of gender disparity in the UK from 2012 to 2017. 

\begin{figure}[ht]
    \centering
     \caption{Malaysian researchers in Computer Science}
    \includegraphics[width=0.5\textwidth]{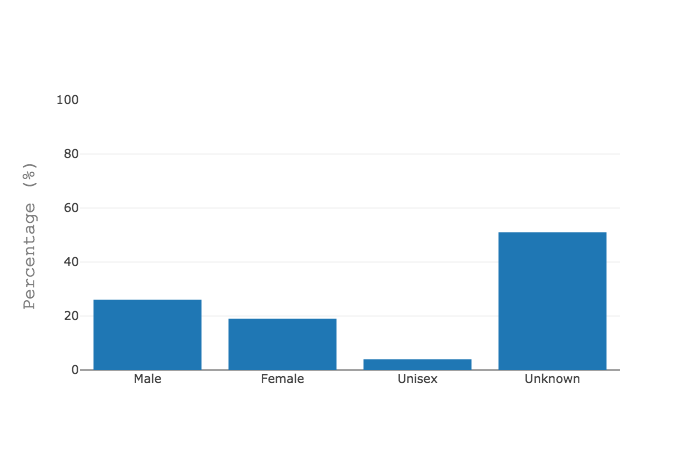}
    \label{fig:MA}
\end{figure}

In Malaysia, there are more male than female computing researchers. Figure \ref{fig:MA} shows the result of the situation on gender disparity in Malaysia from 2012 to 2017. 

\section{Conclusion and Future Work}

In this paper, we have presented a method for analysing online data for the  gender disparity in the computer science field in the UK, Malaysia and China. We improved a gender prediction tool of first names which helps us to complete the online data more accurately in two different languages' characters. The system can display the result to users directly on dynamic graphs. This method is useful for social researchers to process big data when making the gender prediction of first names. We did the experiments with our tool in analysing the gender disparity in computer science in the UK, Malaysia and China. However, we think it is limiting that researching the gender gap in Science depends on this method. There are massive online data that need to be processed as the social research in analysing it. Therefore, we want to develop a new method that can output high accuracy results for predicting gender, data's subjects and their culture origin simultaneously.


\begin{thebibliography}{00}
\bibitem{b1} CNKI.NET. Journal of China Academic Database. Available at: https://www.ssa.gov/oact/babynames.html, Last accessed: June 2017.
\bibitem{b2} Jingchao Hu. ngender 0.1.1: Guess gender for Chinese names. Available at: https://pypi.python.org/pypi/ngender/ 0.1.1, Last accessed: February 2017.
\bibitem{b3} Global gender disparities in science. Vol. 504. Nature, Dec 2013.
\bibitem{b4} MIT. Plotly Python Library. Available at: https://plot.ly/python, Last accessed: August 2017.
\bibitem{b5} Namsor. NamSor Gender API. Available at: http://www. namsor.com, Last accessed: May 2017.
\bibitem{b6} Jacob Perkins. Python Text Processing with NLTK 2.0 Cook- book. Packt Publishing, 9 Nov. 2010. isbn: 1849513600.
\bibitem{b7} python.org. pickle,Python object serialization. Available at: https://docs.python.org/2/library/pickle.html, Last accessed: June 2017.
\bibitem{b8} saedsayad.com. Naive Bayesian. Available at: http://www. saedsayad.com/naivebayesian.htm, Last accessed: May 2017.
\bibitem{b9} U.S.A Social Security. National Data. Available at: https: / / www . ssa . gov / oact / babynames . html, Last accessed: June 2017.
\bibitem{b10} spirit. guess language spirit 0.5.3. Available at: https:// pypi.python.org/pypi/guesslanguage-spirit, Last accessed: June 2017.
\bibitem{b11} textgain.com. TEXTGAIN. Available at: https : / / www . textgain.com, Last accessed: May 2017.
\bibitem{b12} Kamil Wais. Gender Prediction Methods Based on First Names with genderizeR. In: The R Journal 8.1 (2016), pp. 17,37.
\bibitem{b13} webofknowledge.com. Web of Science. Available at: https: //apps.webofknowledge.com, Last accessed: June 2017.
\bibitem{b14} gender-api.com.Gender API. Available at:https://gender-api.com, Last accessed: September 2017.
\bibitem{b15}  Andrew  Flowers.  “The  Most  Common  Unisex  Names  In
America: Is Yours One Of Them?” In:FiveThirtyEight (2015).
\bibitem{b16}  saedsayad.com.Naive Bayesian. Available at: http://www.saedsayad.com/naivebayesian.html, Last accessed: May2017.
\bibitem{b17}  wikipedia.org.List of British scientists. Available at: https://en.wikipedia.org/wiki/ListofBritishscientists, Last accessed: June2017.
\bibitem{b18}  baidu.com. List of Scientist. Available at: http://baike.baidu.com/renwu, Last accessed: June2017.
\end{thebibliography}
\end{document}